\documentclass[twocolumn,showpacs,preprintnumbers,amsmath,amssymb]{revtex4}
\usepackage{graphicx}
\usepackage{dcolumn}
\usepackage{bm}

\begin{document}


\title{Universal role of correlation entropy in critical phenomena}
\author{Shi-Jian Gu$^1$}
\altaffiliation{Email: sjgu@phy.cuhk.edu.hk\\URL: http://www.phystar.net/}
\author{Chang-Pu Sun$^{1, 2}$}
\author{Hai-Qing Lin$^1$} \affiliation{$^1$Department of Physics and ITP, The
Chinese University of Hong Kong, Hong Kong, China \\
$^2$Institute of Theoretical Physics and Interdisciplinary Center of
Theoretical Studies, Chinese Academy of Sciences, Beijing, 100080, China}

\begin{abstract}
In statistical physics, if we successively divide an equilibrium system into
two parts, we will face a situation that, within a certain length $\xi$, the
physics of a subsystem is no longer the same as the original system. Then the
extensive properties of the thermal entropy $S($AB$)= S($A$)+S($B$)$ is
violated. This observation motivates us to introduce the concept of correlation
entropy between two points, as measured by mutual information in the
information theory, to study the critical phenomena. A rigorous relation is
established to display some drastic features of the non-vanishing correlation
entropy of the subsystem formed by any two distant particles with long-range
correlation. This relation actually indicates the universal role of the
correlation entropy in understanding critical phenomena. We also verify these
analytical studies in terms of two well-studied models for both the thermal and
quantum phase transitions: two-dimensional Ising model and one-dimensional
transverse field Ising model. Therefore, the correlation entropy provides us
with a new physical intuition in critical phenomena from the point of view of
the information theory.
\end{abstract}

 \maketitle






\maketitle

\section{Introduction}
Entropy is one of the most important concept in both statistical physics
\cite{Khuangb} and information theory \cite{Shannon48,TMCover91b,Nielsenb}. It
measures how much uncertainty there is in a state of physical system. In
statistical physics, the entropy defined by $S=\log_2 \Omega$ depends on the
number of states $\Omega$ in an equilibrium system; while in the information
theory, the entropy $S=-{\rm tr}(\rho\log_2\rho)$ is associated with the
probability distribution in the eigenstate space of the density matrix $\rho$.
Therefore, it is interesting to look into some fundamental issues from
different point of view due to the common ground of two fields.

In statistical physics, if we divide an equilibrium system into a large number
of macroscopic parts, the total number of state in phase space is a product
form from the number $\omega_i$ of state in each part, i.e.
$\Omega=\prod_i\omega_i$. This precondition leads to that the entropy is an
extensive quantity in statistical physics, i.e. $S=\sum_i S_i$, which is one of
basis of the second law of the thermodynamics. However, a realistic system
includes all kinds of interaction, and the dynamics at one site is no longer
independent of other sites nearby. The facts implies that if we successively
divide a system into two parts, we will face a situation that, within a certain
length $\xi$, the physics of a subsystem is not the same as the original
system. This length actually defines a characteristic scale for the statistical
physics, and physicists usually call it correlation length in the studying of
different kinds of correlation function. From the point view of the information
theory, for two subsystems A and B within the length scale $\xi$, they are no
longer independent with each other, and their entropy does not satisfy the
extensive property, i.e. $S($AB$)\neq S($A$)+S($B$)$.

On the other hand, correlation function plays a fundamental role in physics.
Almost all physical quantities, not only in condensed matter physics, but also
in quantum field theory, are related to the correlation function either
directly or indirectly. Thus it devotes to understand many phenomena in both
classical and quantum mechanics. Physically, the correlation function denotes
the amplitude of the dependence of physical variables between two points in
space time. From the information theory, it also partially measures how much
uncertainty of a physical quantity at one location if the quantity at another
one is given. However, this uncertainty still depends on the quantity itself.
For example, in quantum system, the diagonal correlation function usually
differs from the off-diagonal correlation function. Therefore, in order to
learn the dependence between two separated systems, it is important to ask such
a question: to what extent that the equality $S($AB$)=S($A$)+S($B$)$ is
violated. This question introduces a concept in the information theory, i.e.
mutual information, which is defined as
\begin{eqnarray}
S({\rm A: B})=S({\rm A})+S({\rm B})-S({\rm AB})
\end{eqnarray}
where $S(i)=-{\rm tr}(\rho_i \log_2\rho_i), i={\rm A,B, AB}$ is the entropy of
the corresponding reduced density matrix. If $\rho$ is classical, it is the
Shannon entropy \cite{Shannon48}, otherwise it is the von Neumann entropy
\cite{AWehrl78} of the quantum information theory. Actually, all correlation
functions $\langle O_{\rm A} O_{\rm B}\rangle$ between A and B can be
calculated from the reduced density matrix $\rho_{\rm AB}$, i.e. $\langle
O_{\rm A} O_{\rm B}\rangle={\rm tr}\langle O_{\rm A} O_{\rm B} \rho_{\rm AB}
\rangle$. The mutual information, which measures the common shared information,
defines a more general operator-independent correlation between two subsystems.
Taking into account the role of entropy in the statistical physics, we would
like to call it correlation entropy hereafter.

To have a concrete interpretation, it is useful return to intuitive
understanding of the entropy itself. In the information theory, the entropy is
used to quantify the physical resource (in unit of classical bit due to
$\log_2$ in its expression) needed to store information. For example, in the
exact diagonalization approach, if we want to diagonalize a Hamiltonian of
10-site spin-1 chain and no symmetry can be used to reduce the dimension of the
Hilbert space, we need at least $S=\log_2 3^{10}=10\log_2 3$ bits to store a
basis. Therefore, the correlation entropy actually measures the additional
physical resource required if we store two subsystems respectively rather than
store them together. As a simple example, let us consider a two-qubit system in
a singlet state $(|\uparrow \downarrow\rangle - |\downarrow \uparrow
\rangle)/\sqrt{2}$, we have $S($A$)=1, S($B$)=1, S($AB$)=0$, which leads to
$S($A: B$)=2$. Obviously, there is no information in a given singlet state.
However, each spin in this state is completely uncertain. So we need two bits
to store them respectively. On the other hand, the correlation entropy is
simply the twice of the entanglement, as measured by partial entropy,  between
two systems because of $S($A$)=S($B$)$ for pure state. The reason that we
interpret the correlation entropy in this way is that besides the quantum
correlation the state also has classical correlation. From this point of view,
the correlation entropy is just a measure of total correlation, including
quantum and classical correlation, between two subsystems. They go halves with
each other in correlation entropy for pure state. For a mixed state, the
correlation entropy also measures the amount of the uncertainty of one system
before we learn one from another. From the above interpretation for pure state,
it is then not surprising that the correlation entropy fails to measure the
entanglement \cite{VVedral1997}.

In the statistical physics, the critical phenomena is the central topic. To
have a complete understanding on the critical behavior, various methods, such
as renormalization group \cite{Rshankar94}, Monte-Carlo simulation
\cite{DPLandaub}, and mean-field approach etc., have been developed and applied
to many kinds of systems. In recent years, the study on the role of
entanglement in the quantum critical behavior \cite{Sachdev} have established a
bridge between quantum information theory and condensed matter physics, and
shed new light on the quantum phase transitions due to its interesting behavior
around the critical point
\cite{AOsterloh2002,TJOsbornee,SJGuXXZ,SJGUPRL,Aanfossi05,HTQuan06}. However,
the entanglement is fragile under the thermal fluctuation and can be suppressed
to zero at finite temperatures. Then it is difficult to witness a generalized
thermal phase transition in terms of quantum entanglement.

In this paper, we are going to study the role of correlation entropy in both
thermal and quantum phase transitions. Like the fundamental role of two point
correlation function in the statistical physics, we are interested in the
universal role of two-point correlation entropy in the present work. The paper
is organized as follows. In Sec. \ref{sec:toymodel}, for the pedagogical
purpose, we study a toy model and show that the thermal entropy is not an
extensive parameter in such a simple system. In Sec. \ref{sec:order}, we
discuss the relation between the reduced density matrix, long-range
correlation, and the correlation entropy. In Sec. \ref{sec:2disingmodel}, we
study the properties of the correlation entropy in thermal phase transition, as
illustrated by classical two-dimensional Ising model. In Sec.
\ref{sec:Isingmodel}, we address the properties of the correlation entropy in a
simple quantum phase transition of one-dimensional transverse field Ising
model. In Sec. \ref{sec:dis}, some discussions and prospects are presented.
Finally, a brief summary is given in Sec. \ref{sec:sum}

\section{Toy model: Heisenberg dimer}
\label{sec:toymodel}

For the pedagogical purpose and making our motivation more clear, we first have
a look on a very simple model: a Heisenberg dimer. Its Hamiltonian reads
\begin{eqnarray}
H=\sigma_1\cdot\sigma_2
\end{eqnarray}
where $\sigma_i (\sigma^x, \sigma^y, \sigma^z)$ are Pauli matrices at site $i$,
\begin{eqnarray}
\sigma^x=\left( %
\begin{array}{cc}
  0 & 1 \\
  1 & 0 \\
\end{array}%
\right),\; \sigma^y=\left(%
\begin{array}{cc}
  0 & -i \\
  i & 0 \\
\end{array}%
\right),\; \sigma^z=\left(%
\begin{array}{cc}
  1 & 0 \\
  0 & -1 \\
\end{array}%
\right)
\end{eqnarray}
and the coupling between two sites is set to unit for simplicity. The
Hamiltonian can be diagonalized easily. Its ground state is a spin singlet
state
\begin{eqnarray}
\Psi_0=\frac{1}{\sqrt{2}}\left[|\uparrow\downarrow
\rangle-|\downarrow\uparrow\rangle\right],
\end{eqnarray}
with eigenvalue $E_0=-3$, while three degenerate excited states are
\begin{eqnarray}
\Psi_1=\frac{1}{\sqrt{2}}\left[|\uparrow\downarrow \rangle
+|\downarrow\uparrow\rangle\right],\nonumber \\
\Psi_2=|\uparrow\uparrow\rangle,\;\;\; \Psi_3=|\downarrow\downarrow\rangle.
\end{eqnarray}
with higher eigenvalue $E_{1,2,3}=1$. Therefore, according to the statistical
physics, the thermal entropy of the system vanishes at zero temperature. When
the system is contacted with a thermal bath with temperature $T$, the thermal
state of the system is described by a density matrix:
\begin{eqnarray}
\rho=\zeta\left(
                                  \begin{array}{cccc}
                                    e^{-2/T} & 0 & 0 & 0 \\
                                    0 & \cosh(2/T) & -\sinh(2/T) & 0 \\
                                    0 & \sinh(2/T) & \cosh(2/T) & 0 \\
                                    0 & 0 & 0 & e^{-2/T} \\
                                  \end{array}
                                \right)
\end{eqnarray}
where
\begin{eqnarray}
\zeta=\frac{1}{3e^{-2/T}+e^{2/T}}.
\end{eqnarray}
Because of SU(2) symmetry of the Heisenberg dimer, the single-site entropy
$S(i)$ of the system is always unity, and the entropy of the whole system is
\begin{eqnarray}
S(12)=2 -\frac{\log_2(3e^{-2/T}+1)}{3e^{-2/T}+1}
-\frac{3\log_2(3+e^{4/T})}{3+e^{4/T}}.
\end{eqnarray}
Then we can see that $S(12)\neq S(1)+S(2)$ both at zero temperature and finite
temperatures. In high temperature limit, the asymptotic behavior of the
correlation entropy is $S(1: 2)\propto 1/T$. So only when $T\rightarrow
\infty$, $S(1: 2)\rightarrow 0$, the extensive property of the entropy holds.
The physics behind this fact is quite clear. It is the interaction between two
sites that establishes a kind of correlation and then breaks the extensive
property of the entropy.

For a large system, however, this correlation usually decays with the
increasing of distance between two parts, and the entropy becomes an extensive
quantity beyond a definite scale. Only around the critical point where the
phase transition happens, the system behaves like a whole and can not be
divided into two part, and then the correlation entropy has long range
behaviors.

\section{reduced density matrix, long-range order, and correlation entropy}
\label{sec:order}

In many-body physics, the reduced density matrix of a one-body and two-body
subsystem can be, in general, written as
\begin{eqnarray}
&&\langle\alpha'|\rho_i|\alpha\rangle={\rm tr}(a_{i\alpha'}\rho
a^\dagger_{i\alpha}),\nonumber \\
&&\langle\alpha'\beta'|\rho_{ij}|\alpha\beta\rangle={\rm
tr}(a_{i\alpha'}a_{j\beta'}\rho a^\dagger_{j\beta}
a^\dagger_{i\alpha}),\label{eq:DefRDM}
\end{eqnarray}
respectively. Here $a_{i\alpha}, a_{j\beta}$ are annihilation operators for
states $|\alpha\rangle, |\beta\rangle$ localized at site $i, j$ respectively,
and satisfy commutation (anti-commutation) relation for bosonic (fermionic)
states. The reduced density matrix is usually normalized as
\begin{eqnarray}
{\rm tr}(\rho_i)=1,\;\; {\rm tr}(\rho_{ij})=1, \label{eq:rdmnormcond}
\end{eqnarray}
so that one has a probability explanation for their diagonal elements in the
corresponding eigenstate space.

In spin system, the reduced density matrix of a single spin at position $i$
takes the form
\begin{eqnarray}
\rho_i=\frac{1}{2}\big(1+\langle\sigma_i^x\rangle\sigma_i^x
+\langle\sigma_i^y\rangle\sigma_i^y +\langle\sigma_i^z\rangle\sigma_i^z \big).
\end{eqnarray}
For two arbitrary spins at position $i$ and $j$, the two-site reduced density
matrix generally takes the form,
\begin{eqnarray}
\rho_{ij}&=&\frac{1}{4}+\frac{1}{4}\sum_\alpha\left(
\langle\sigma_i^\alpha\rangle\sigma_i^\alpha
+\langle\sigma_j^\alpha\rangle\sigma_j^\alpha\right)\nonumber
\\ &&+\frac{1}{4}\sum_{\alpha\beta}
\langle\sigma_i^\alpha\sigma_j^\beta\rangle\sigma_i^\alpha\sigma_j^\beta.
\label{eq:spinrdm_def}
\end{eqnarray}
Obviously, under some symmetry, the above reduced density matrix can be
simplified. For example, if the state of $N$ spins is also an eigenstate of $z$
component of total spins $S^z=\sum S_i^z=0$ and possesses the exchange
symmetry, then Eq. (\ref{eq:spinrdm_def}) can be simplified as
\begin{eqnarray}
\rho_{ij}=\left(%
\begin{array}{cccc}
  u^+ & 0 & 0 & 0 \\
  0 & w_1 & z & 0 \\
  0 & z^* & w_2 & 0 \\
  0 & 0 & 0 & u^-
\end{array}%
\right)
\end{eqnarray}
in the basis of $\sigma^z_i\sigma^z_j$: $\{|\uparrow\uparrow\rangle$,
$|\uparrow\downarrow\rangle$, $|\downarrow\uparrow\rangle$,
$|\downarrow\downarrow\rangle \}$. Here the matrix elements can be calculated
from the correlation function,
\begin{eqnarray}
&&u^+=u^-=\frac{1}{4}(1+\langle\sigma^z_i\sigma^z_j\rangle), \nonumber \\
&&w_1=w_2=\frac{1}{4}(1-\langle\sigma^z_i\sigma^z_j\rangle), \nonumber \\
&&z=\frac{1}{4}(\langle\sigma^x_i\sigma^x_j\rangle+\langle\sigma^y_i\sigma^y_j\rangle).
\label{eq:rdmelements1}
\end{eqnarray}

With the help of the Jordan-Schwinger mapping \cite{JordanSchingerm},
\begin{eqnarray}
&&\sigma _{j}^{+}=a_{j\uparrow }^{\dagger }a_{j\downarrow },\nonumber \\
&&\sigma_{j}^{-}=a_{j\downarrow }^{\dagger }a_{j\uparrow },\nonumber \\
&&\sigma_{j}^{z}=(a_{j\uparrow }^{\dagger }a_{j\uparrow }-a_{j\downarrow
}^{\dagger }a_{j\downarrow })
\end{eqnarray}
where $a_{j\alpha}^\dagger$ stands for the pseudo fermionic creation operator
for single particle state $|j,\alpha \rangle $ at the position $j$, the element
in the reduced density matrix (\ref{eq:spinrdm_def}) can be reexpressed in the
form of Eq. (\ref{eq:DefRDM}). For example,
\begin{eqnarray}
\langle \sigma _{i}^{+}\sigma _{j}^{-}\rangle =-\langle a_{i\uparrow }^{\dagger
}a_{j\downarrow }^{\dagger }a_{i\downarrow }a_{j\uparrow }\rangle.
\end{eqnarray}
Therefore, we can explore the property of long-range correlation in spin system
through pseudo fermion systems.

We first consider the long-range correlation in classical systems, e.g. Ising
model, in which the reduced density matrix takes the diagonal form, i.e.
\begin{eqnarray}
&&\langle\alpha'|\rho_i|\alpha\rangle=\delta_{\alpha'\alpha}{\rm
tr}(a_{i\alpha'}\rho
a^\dagger_{i\alpha}),\nonumber \\
&&\langle\alpha'\beta'|\rho_{ij}|\alpha\beta\rangle=\delta_{\alpha'\alpha}\delta_{\beta'\beta}
{\rm tr}(a_{i\alpha'}a_{j\beta'}\rho a^\dagger_{j\beta} a^\dagger_{i\alpha}).
\end{eqnarray}
Then, if there is no long-range correlation
\begin{eqnarray}
\langle\alpha\beta|\rho_{ij}|\alpha\beta\rangle &=& {\rm tr}(a^\dagger_{j\beta}
a^\dagger_{i\alpha} a_{i\alpha}a_{j\beta}\rho ),\nonumber \\
&=& \langle a^\dagger_{j\beta} a^\dagger_{i\alpha} a_{i\alpha}a_{j\beta}\rangle
,\nonumber \\
&\rightarrow& \langle a^\dagger_{i\alpha} a_{i\alpha}\rangle \langle
a^\dagger_{j\beta} a_{j\beta}\rangle,
\end{eqnarray}
for ${|i-j|\rightarrow \infty}$, the two-site entropy becomes
\begin{eqnarray}
S(ij) \rightarrow S(i)+S(j),
\end{eqnarray}
where
\begin{eqnarray}
S(l)= - \sum_\beta \langle a^\dagger_{l\beta} a_{l\beta}\rangle \log_2 \langle
a^\dagger_{l\beta} a_{l\beta}\rangle,\;\;l=i, j
\end{eqnarray}
where the normalization conditions of the $\rho_{i(j)}$ and $\rho_{ij}$ have
been used. Obviously, we have $S(i:j)=0$, which means that if there is no
long-range correlation, the correlation entropy vanishes at long distance.

On the other hand, if there exists long-range correlation, for example
\begin{eqnarray}
G_{i\alpha,\,j\beta} \equiv  \langle n_{i\alpha} n_{j\beta}\rangle - \langle
n_{i\alpha}\rangle\langle n_{j\beta}\rangle = C,
\end{eqnarray}
for ${|i-j|\rightarrow \infty}$ and a constant $C$, then
\begin{eqnarray}
S(i: j) > 0.
\end{eqnarray}

Now we study the correlation entropy in the quantum systems, in which the
reduced density matrix usually is not diagonal. Then, if the system does not
have long-range correlation,
\begin{eqnarray}
\langle\alpha'\beta'|\rho_{ij}|\alpha\beta\rangle= \langle a_{i\alpha}^\dagger
a_{i\alpha'} \rangle \langle a_{j\beta}^\dagger a_{j\beta'} \rangle,
\end{eqnarray}
for ${|i-j|\rightarrow \infty}$, the reduced density matrix can be written into
a direct product form, i.e.
\begin{eqnarray}
\rho_{ij}=\rho_i \otimes \rho_j .
\end{eqnarray}
Then the reduced density matrices $\rho_i, \rho_j$ can be diagonalized in their
own subspace. So in principle, we can have $\rho_i=\sum_\alpha
p_\alpha|\varphi_\alpha\rangle_{ii}\langle\varphi_\alpha|$, and
$\rho_j=\sum_\beta p_\beta|\varphi_\beta\rangle_{jj}\langle\varphi_\beta|$,
where $p_\alpha, p_\beta$ are the probability distribution for $\rho_i$ and
$\rho_j$ respectively. As we have done for the classical system, we then have
\begin{eqnarray}
S(i: j)=S(\rho_i)+S(\rho_j)-S(\rho_{ij})=0.
\end{eqnarray}

In order to study its relation to the long-range correlation, now we express the
correlation entropy in term of the relative entropy \cite{Vvedral02}
\begin{equation}
S(i:j)=\mathrm{tr}(\rho _{ij}\log _{2}\rho _{ij})-\mathrm{tr}(\rho _{ij}\log
_{2}\rho _{i}\otimes \rho _{j})
\end{equation}
between the whole system and direct product form of two subsystems where $(\rho
_{i}\otimes \rho _{j})_{\alpha \beta ,\alpha ^{\prime }\beta ^{\prime }}=\langle
a_{i\alpha }^{\dagger }a_{i\alpha ^{\prime }}\rangle \langle a_{j\beta
}^{\dagger }a_{j\beta ^{\prime }}\rangle $. For Hermition operators $\rho_{ij}$,
the eigen-equation  $\rho _{ij}|\phi _{\alpha \beta }\rangle =q_{\alpha \beta
}|\phi _{\alpha \beta }\rangle $ gives  a representation of the correlation
entropy in the basis  $\{|\phi _{\alpha \beta }\rangle \}$. Therefore, in the
eigenstate space of $|\phi_{\alpha\beta}\rangle$, we have
\begin{eqnarray}
&&S(i: j)=\sum_{\alpha\beta}q_{\alpha\beta}\log_2 q_{\alpha\beta} \nonumber \\
&&\;\;\;\;\;\;\;\;\;- \sum_{\alpha\beta}\langle
\phi_{\alpha\beta}|\rho_{ij}\log_2 (\rho_i \otimes \rho_j)|\phi_{\alpha
\beta}\rangle.
\end{eqnarray}
Insert the identity $\sum_{\alpha\beta}|\varphi_\alpha\varphi_\beta\rangle
\langle\varphi_\beta\varphi_\alpha | =1$, it becomes
\begin{eqnarray}
S(i: j)= \sum_{\alpha\beta}q_{\alpha\beta}\log_2 q_{\alpha\beta} -
\sum_{\alpha\beta\alpha'\beta'}P_{\alpha'\beta',
\alpha\beta}q_{\alpha\beta}\log_2 (p_\alpha p_\beta)\nonumber
\end{eqnarray}
where
\begin{eqnarray}
P_{\alpha'\beta', \alpha\beta}=
\langle\phi_{\alpha\beta}|\varphi_{\alpha'}\varphi_{\beta'}\rangle
\langle\varphi_{\alpha'}\varphi_{\beta'}|\phi_{\alpha\beta}\rangle
\end{eqnarray}
and satisfies
\begin{eqnarray}
 P_{\alpha'\beta', \alpha\beta} >0, \;\; \sum_{\alpha'\beta'}P_{\alpha'\beta', \alpha\beta}=1,
\;\sum_{\alpha\beta}P_{\alpha'\beta', \alpha\beta}=1.
\end{eqnarray}
Moreover, $P_{\alpha'\beta', \alpha\beta}$ will become unit matrix if and only
if $\langle\phi_{\alpha\beta}|\varphi_\alpha\varphi_\beta\rangle=1$.  Then
\begin{eqnarray}
&&S(i: j)
\\ \nonumber =&&\sum_{\alpha\beta}q_{\alpha\beta} \left(\log_2
q_{\alpha\beta}-\sum_{\alpha'\beta'}P_{\alpha'\beta',
\alpha\beta}\log_2(p_{\alpha'}p_{\beta'})\right).
\end{eqnarray}
Taking into account the concavity property of $\log$ function, i.e.
$$\sum_{\alpha'\beta'}P_{\alpha'\beta',
\alpha\beta}\log_2(p_{\alpha'}p_{\beta'})
\leq\log_{2}\left(\sum_{\alpha'\beta'}P_{\alpha'\beta', \alpha\beta}p_\alpha
p_\beta\right),$$  we can have
\begin{eqnarray}
S(i: j) &\geq & \sum_{\alpha\beta}q_{\alpha\beta}\log_2 Q_{\alpha\beta}
\nonumber  \\ &\geq & \frac{1}{\ln 2}\sum_{\alpha\beta}  q_{\alpha\beta}
\left(1- Q_{\alpha\beta}\right)=0,
\end{eqnarray}
where
\begin{eqnarray}
Q_{\alpha\beta}= \frac{q_{\alpha\beta}}{\sum_{\alpha'\beta'}P_{\alpha'\beta',
\alpha\beta}p_\alpha p_\beta},
\end{eqnarray}
The above inequality becomes an equality if and only if $P_{\alpha'\beta',
\alpha\beta}$ is an unit matrix. Therefore, if
\begin{equation}
\langle \alpha ^{\prime }\beta ^{\prime }|\rho _{ij}|\alpha \beta \rangle
=\langle a_{i\alpha }^{\dagger }a_{i\alpha ^{\prime }}\rangle \langle a_{j\beta
}^{\dagger }a_{j\beta ^{\prime }}\rangle +C,
\end{equation}
for ${|i-j|\rightarrow \infty }$, $P$ is not diagonal, then we have the desired
result
\begin{eqnarray}
S(i: j)>0. \label{eq:posvend}
\end{eqnarray}

In the information theory, the inequality similar to the above result is called
Klein inequality \cite{Oklein31}. Therefore the existence of the long-range
correlation will lead to a positive correlation entropy (\ref{eq:posvend}).
This observation is very important in understanding critical phenomena.
According to the theory of either thermal or quantum phase transition, the
presence of the long-range correlation is crucial. However, different phase
transition depends on the different long-range correlation. For example, in the
superfluid phase of $^4$He, the off-diagonal-long-range order, as suggested by
Yang \cite{CNYangODLRO}, is necessary; while in anther kind of condensate of
exciton, it may require diagonal-long-range order \cite{Wkohn70}. Then the
above results show that the non-vanishing positive defined correlation is a
universal and necessary condition for all critical phenomena \cite{Note1}.

\section{Thermal phase transition: two-dimensional Ising model}
\label{sec:2disingmodel}

The physics in the above toy model is quite limited. In order to verify our
analytical result and see the significance of the correlation entropy in the
critical phenomena, let us first study its properties in a thermodynamical
system. One of typical examples is the two-dimensional Ising model, which is
certainly the most thoroughly researched model in statistical physics
\cite{LOnsager44,BMMccoyb}.

In the absence of external field, the model Hamiltonian defined on a square
lattice field reads
\begin{eqnarray}
H=-\sum_{\bf \langle i j \rangle} \sigma_{\bf i}^z \sigma_{\bf j}^z,
\end{eqnarray}
where the sum is over all pairs of nearest-neighbor sites ${\bf i}$ and ${\bf
j}$, and the coupling is set to unit for simplicity. Since the Ising model is a
classical model, the reduced density matrix of two arbitrary sites then takes
the form
\begin{equation}
{\rho}_{\bf ij} = \left(
\begin{array}{llll}
u^+ & 0 & 0 & 0 \\
0 & w_1 & 0 & 0 \\
0 & 0 & w_2 & 0 \\
0 & 0 & 0 & u^-
\end{array}
\right)
\end{equation}
in which the elements can be calculated from Eq. (\ref{eq:rdmelements1}), and
single-site reduced density matrix
\begin{equation}
{\rho}_{\bf i} = \frac{1}{2}\left( \begin{array}{cc}
1+\langle\sigma^z_i\rangle& 0  \\
0 & 1-\langle\sigma^z_i\rangle \\
\end{array}
\right), \label{eq:singlesiterdm}
\end{equation}

For simplicity, we only consider the correlation entropy along (1, 1)
direction, because the long distance behavior of the correlation entropy should
be independent of the direction.

\begin{figure}
\includegraphics[width=8cm]{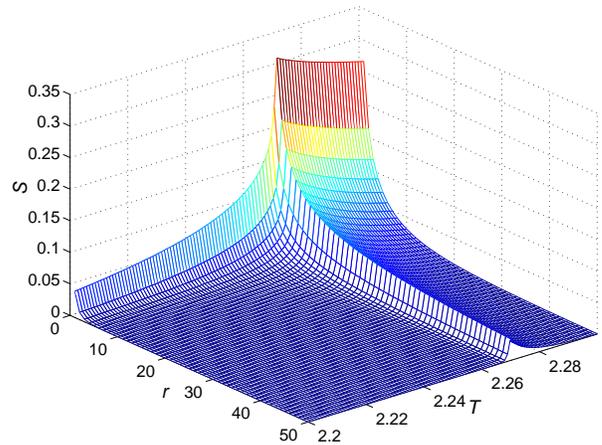}
\caption{\label{figure_2dising}(color online) The correlation entropy as a
function of temperature $T$ (in unit of Ising coupling) and the distance $r$
(in unit of $\sqrt{2}$ lattice constant). }
\end{figure}

According to the exact solution of the two-dimensional Ising model
\cite{BMMccoyb}, the magnetization per site of the system is
\begin{eqnarray}
\langle\sigma^z_{\bf i}\rangle=\left\{
\begin{array}{cc}
\left[1-\sinh^{-4}(2/T)\right]^{1/8} &\;\; T<T_c \\
0 & \;\;T>T_c
\end{array}
\right. \label{eq:Isingmeanmag}
\end{eqnarray}
where the critical temperature $T_c$ is determined by
\begin{eqnarray}
2\tanh(2/T)=1,
\end{eqnarray}
then $T_c \simeq 2.269185$. The correlation function can be calculated as
\begin{eqnarray}
\langle\sigma^z_{0,0}\sigma^z_{N,N}\rangle=\left|\begin{array}{cccc}
                                                   a_0 & a_{-1} & \cdots & a_{-N+1} \\
                                                   a_1 & a_0 & \cdots & a_{-N+2} \\
                                                   \vdots & \vdots & \vdots & \vdots \\
                                                   a_{N-1} & a_{N-2} & \cdots &
                                                   a_0
                                                 \end{array}
\right|
\end{eqnarray}
where
\begin{eqnarray}
a_n=\frac{1}{2\pi}\int_0^{2\pi}d\theta e^{in\theta}\phi(\theta),
\end{eqnarray}
and
\begin{eqnarray}
\phi(\theta)=\left[
\frac{\sinh^2(2/T)-e^{-i\theta}}{\sinh^2(2/T)-e^{i\theta}}\right]^{1/2}
\end{eqnarray}
Therefore, we can calculate the correlation entropy directly from the known
results. We show the correlation entropy $S({\bf i: j})$ as a function of
temperature $T$ and distance between two site ${\bf r=i-j}$ in Fig.
\ref{figure_2dising}.

\begin{figure}
\includegraphics[width=8.1cm]{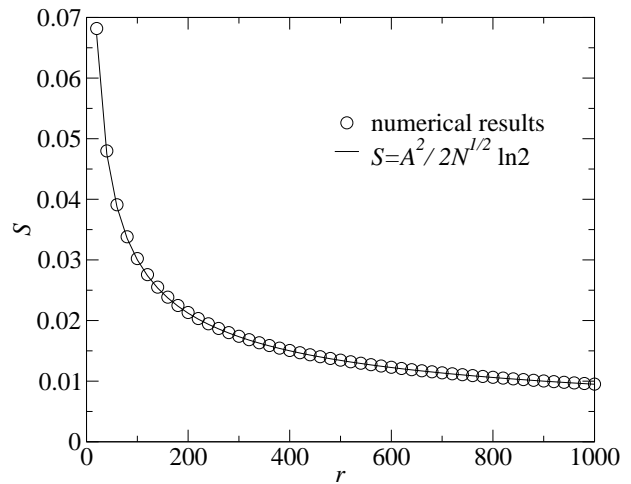}
\caption{\label{figure_2discri_ent}(color online) The correlation entropy as a
function of the distance $r$ (in unit of $\sqrt{2}$ lattice constant) at the
critical point.}
\end{figure}

\begin{figure}[tbp]
\includegraphics[width=8cm]{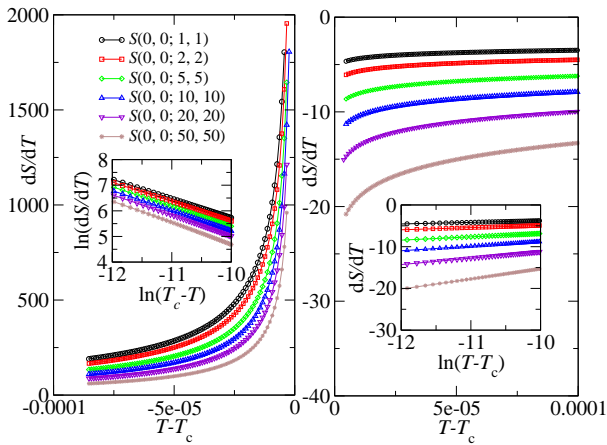}
\caption{(color online) The critical behavior of the correlation entropy. LEFT:
$dS({0,0: N, N})/dT$ below $T_c$, and the inset is to explore its critical
exponent. RIGHT: $dS({0,0: N, N})/dT$ above $T_c$. } \label{figure_2dcri_exp}
\end{figure}

The result is impressive. It is well known that the two-dimensional Ising model
has two different phases separated by $T_c$. Below $T_c$, the system has
macroscopic magnetization, i.e., spontaneously magnetized, and its mean
magnetization is determined by Eq. (\ref{eq:Isingmeanmag}). While above $T_c$,
the thermal fluctuation destroy this order and the system becomes paramagnetic.
Therefore, it is not difficult to understand that the correlation entropy
between two sites decay quickly as the distance increases. This fact implies
that the extensive property of the entropy holds beyond a finite correlation
length. So the physics in a small system can be used to described that for a
large system. It is also the reason why in the Monte Carlo approach,  a
simulation on a small system at low temperature and higher temperature agree
with the analytic result in thermodynamic limit excellently. However, in the
critical region, as we can see from Fig. \ref{figure_2dising}, the correlation
entropy decays in a power-law way. This fact not only tells us a strong
dependence between arbitrary two sites in the system, but also manifests the
integrality of the whole system. It is also the reason of the difficulty of the
Monte-Carlo simulation around the critical region.

Moreover, since the correlation entropy comprises all kinds of two-point
correlation function in its expression, it can tell us more details about the
critical phenomena. At the critical point, $S({\bf i})=1$ because of
$\langle\sigma^z_{\bf i}\rangle=0$, and the correlation function behaviors like
\begin{eqnarray}
S_N\equiv\langle \sigma _{0,0}^{z}\sigma _{N,N}^{z}\rangle \simeq A/N^{1/4},
\end{eqnarray}
where $A\simeq 0.645$. Then the two-site entropy can be simplified as
\begin{eqnarray}
S({\bf ij})&=&2-\frac{1}{2}\left[(1+S_N)\log_2(1+S_N)\right. \nonumber
\\ &&\left.+(1-S_N)\log_2(1-S_N) \right],
\end{eqnarray}
in the large $N$ limit, and the correlation entropy behaviors like
\begin{eqnarray}
S(0,0: N, N)=\frac{A^2}{2N^{1/2}\ln 2},
\end{eqnarray}
as has been shown in Fig. \ref{figure_2discri_ent}. On the other hand,  around
the critical point, the correlation entropy can be written as
\begin{eqnarray}
S({\bf i: j})\simeq \frac{1}{2}\langle\sigma^z_{\bf i}\sigma^z_{\bf j}\rangle^2-
\langle\sigma^z_{\bf i}\sigma^z_{\bf j}\rangle \langle\sigma^z_{\bf i}\rangle^2
\end{eqnarray}
Then
\begin{eqnarray}
\frac{\partial S({\bf i: j})}{\partial T} &=& \left[ \langle\sigma^z_{\bf
i}\sigma^z_{\bf j}\rangle-\langle\sigma^z_{\bf i}\rangle^2\right]\frac{\partial
\langle\sigma^z_{\bf i}\sigma^z_{\bf j}\rangle}{\partial T} \nonumber \\ &&-
2\langle\sigma^z_{\bf i}\sigma^z_{\bf j}\rangle \langle\sigma^z_{\bf
i}\rangle\frac{\partial \langle\sigma^z_{\bf i}\rangle}{\partial T}
\end{eqnarray}
Therefore, in the critical region below $T_c$, the dominant term in $\partial
S({\bf i, j})/\partial T$ is $2\langle\sigma^z_{\bf i}\sigma^z_{\bf j}\rangle
\langle\sigma^z_{\bf i}\rangle{\partial \langle\sigma_{\bf i}\rangle}/{\partial
T}$, which leads to that $\partial S({\bf i, j})/\partial T$ diverges as
$T\rightarrow T_c$, and scales like
\begin{eqnarray}
\frac{\partial S({\bf i: j})}{\partial T} \propto |T-T_c|^{-3/4},
\end{eqnarray}
as we can see from the left picture in Fig. \ref{figure_2dcri_exp}. Then the
critical exponents of $\partial S({\bf i, j})/\partial T$ below $T_c$ is 3/4,
which is consistent with the critical exponent 1/8 of $\langle\sigma_{\bf
i}\rangle$. While in the critical region above $T_c$, $\langle\sigma^z_{\bf
i}\rangle$ vanishes and the dominating term in $S({\bf i: j})$ becomes the
correlation function. Then $\partial S({\bf i, j})/\partial T$ scales like
\begin{eqnarray}
\frac{\partial S({\bf i: j})}{\partial T} \propto \ln|T-T_c|,
\end{eqnarray}
which is the same as the specific heat $C_v$. Therefore, the critical exponent
now becomes 0 (See the right picture in Fig. \ref{figure_2discri_ent} ).
Moreover, we also note that the slope of lines in the right inset of Fig.
\ref{figure_2discri_ent} is not the same. This is due to the fact that the
exponent of the correlation function $\nu$ introduce the distance dependence in
the $\partial S({\bf i: j})/\partial T$ above $T_c$.

\section{Quantum phase transition: one-dimensional transverse field Ising model}
\label{sec:Isingmodel}

We now study the correlation entropy in one-dimensional transverse field Ising
model whose Hamiltonian reads
\begin{eqnarray}
&&H_{\rm Ising
}=-\sum_{j=1}^N\left[\lambda\sigma_j^x\,\sigma_{j+1}^x+\sigma_j^z\right],\nonumber
\\ && \sigma_1=\sigma_{N+1} \label{eq:hamitl2}
\end{eqnarray}
where $\lambda$ is an Ising coupling in unit of the transverse field. The
Hamiltonian changes the number of down spins by two, the total space of system
then can be divided by the parity of the number of down spins. That is the
Hamiltonian and the parity operator $P=\prod_j\sigma_j^z$ can be simultaneously
diagonalized and the eigenvalues of $P$ is $\pm 1$. We confine our interesting
to the correlation entropy between two spins at position $i$ and $j$ in the
chain. Therefore we need to consider both single-site reduced density matrix
$\rho_i$ obtained from the ground-state wave function by tracing out all spins
except the one at site $i$, and the two-site reduced density matrix $\rho_{ij}$
obtained by tracing out all spins except those at site $i$ and $j$. Then if
there is no symmetry broken, such as in a finite-size system, according to the
parity conservation, $\rho_i$ has a diagonal form Eq. (\ref{eq:singlesiterdm}),
and the reduced density matrix of two spins on a pair of lattice sites $i$ and
$j$ can be put into the following block-diagonal form
\begin{equation}
{\rho}_{ij} = \left(
\begin{array}{llll}
u^+ & 0 & 0 & z^- \\
0 & w_1 & z^+ & 0 \\
0 & z^+ & w_2 & 0 \\
z^- & 0 & 0 & u^-
\end{array}
\right)
\end{equation}
in the basis $|\uparrow\uparrow\rangle, |\uparrow\downarrow\rangle,
|\downarrow\uparrow\rangle, |\downarrow\downarrow\rangle$. The elements in the
density matrix $\rho_{ij}$ can be calculated from the correlation function.
\begin{eqnarray}
&&u^\pm=\frac{1}{4}(1\pm2\langle\sigma_i^z \rangle
+\langle\sigma_i^z\sigma_j^z\rangle), \nonumber \\
&&w_1=w_2=\frac{1}{4}(1- \langle\sigma_i^z\sigma_j^z\rangle), \nonumber \\
&&z^\pm = \frac{1}{4}(\langle\sigma_i^x\sigma_j^x\rangle
\pm\langle\sigma_i^y\sigma_j^y\rangle)
\end{eqnarray}
Otherwise, if the symmetry is broken at the ground state of the ordered phase
in the thermodynamic limit, i.e. $\langle\sigma^x\rangle\neq 0$, then the
single-site reduced density matrix becomes
\begin{eqnarray}
{\rho}_{i} = \frac{1}{2}\left( \begin{array}{cc}
1+\langle\sigma^z_i\rangle& \langle\sigma_i^x\rangle  \\
\langle\sigma_i^x\rangle & 1-\langle\sigma^z_i\rangle \\
\end{array}
\right),
\end{eqnarray}
and the two-site reduced density matrix return to the original form
(\ref{eq:spinrdm_def}) since no symmetry can be used to simply it. Therefore,
the correlation entropy between two sites $i$ and $j$ becomes
\begin{eqnarray}
S(i:j)=2{\rm tr}\rho_i\log_2\rho_i - {\rm tr}\rho_{ij}\log_2\rho_{ij}.
\end{eqnarray}
Taking into account the translation invariance, the correlation entropy is
simply a function of the distance between two sites.

\begin{figure}
\includegraphics[width=8cm]{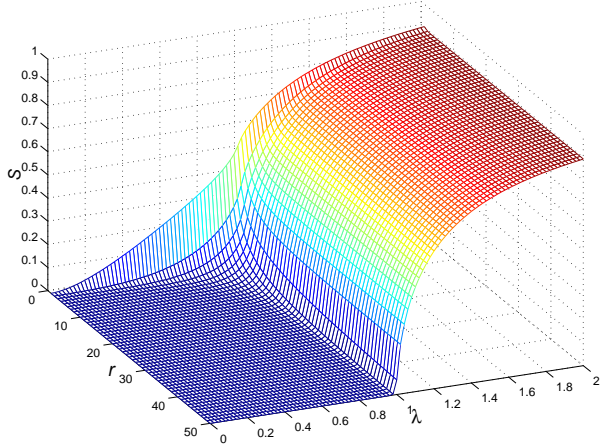}
\caption{\label{figure_centropy_00}(color online) The correlation entropy as a
function of $\lambda$ and the distance $r$ (in unit of lattice constant) at
$T=0$ for a system with $N=5000$.}
\end{figure}

\begin{figure}
\includegraphics[width=8cm]{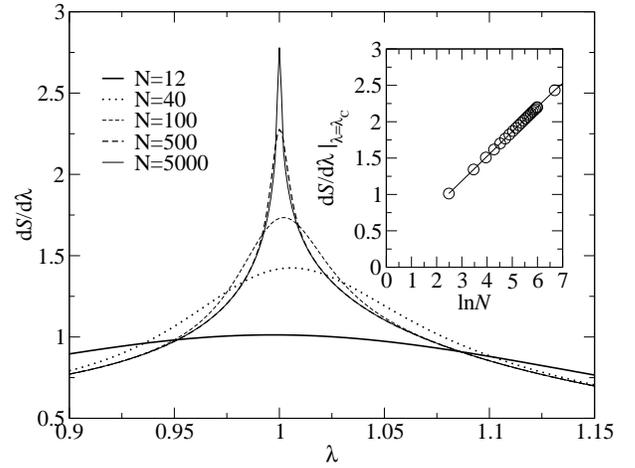}
\caption{The scaling behavior the correlation entropy between two neighboring
sites. } \label{figure_centropy_scaling}
\end{figure}

\begin{figure}
\includegraphics[width=8cm]{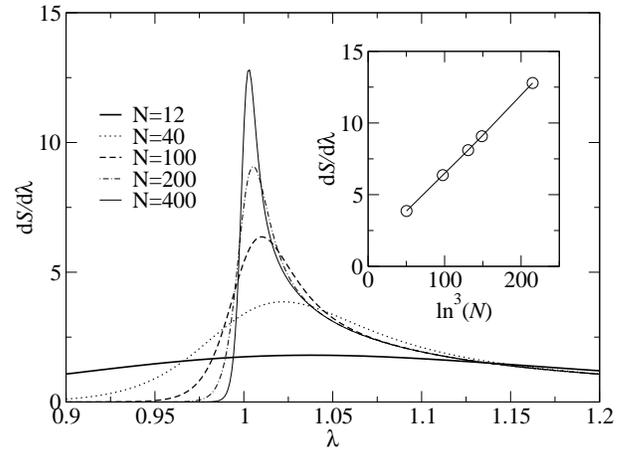}
\caption{The scaling behavior the correlation entropy between two sites at the
longest distance. } \label{figure_centropy_scaling_long}
\end{figure}

The transverse field Ising model can be solved exactly in terms of Jordan-Wigner
transformation. The mean magnetization is given by \cite{EBarouch70}
\begin{eqnarray}
\langle\sigma^z\rangle = \frac{1}{N}\sum_\phi \frac{(1-\lambda\cos\phi)
\tanh[\omega_\phi/T]}{\omega_\phi},
\end{eqnarray}
where $\omega_\phi$ is the dispersion relation,
\begin{eqnarray}
&&\omega_\phi=\sqrt{1+\lambda^2-2\lambda\cos(\phi_q)},\nonumber \\ &&\phi_q=2\pi
q/N,
\end{eqnarray}
where $q$ is integer (half-odd integer) for parity $P=-1 (+1)$. The two-point
correlation functions are calculated as \cite{EBarouch71}
\begin{eqnarray}
\langle\sigma_0^x\sigma_N^x\rangle=\left|\begin{array}{cccc}
                                                   a_{-1} & a_{-2} & \cdots & a_{-N} \\
                                                   a_0 & a_{-1} & \cdots & a_{-N+1} \\
                                                   \vdots & \vdots & \vdots & \vdots \\
                                                   a_{N-2} & a_{N-3} & \cdots &
                                                   a_{-1}
                                                 \end{array}
\right|
\end{eqnarray}
\begin{eqnarray}
\langle\sigma_0^y\sigma_N^y\rangle=\left|\begin{array}{cccc}
                                                   a_{1} & a_{0} & \cdots & a_{-N+2} \\
                                                   a_2 & a_{1} & \cdots & a_{-N+3} \\
                                                   \vdots & \vdots & \vdots & \vdots \\
                                                   a_{N} & a_{N-1} & \cdots &
                                                   a_{1}
                                                 \end{array}
\right|
\end{eqnarray}
\begin{eqnarray}
\langle\sigma_0^z\sigma_N^z\rangle = 4\langle\sigma^z\rangle^2 - a_N a_{-N}
\end{eqnarray}
where
\begin{eqnarray}
a_N=\frac{1}{N}\sum_\phi \frac{\cos(\phi N)(\lambda\cos\phi -1)
\tanh[\omega_\phi/T]}{\omega_\phi}\nonumber \\
-\frac{\lambda}{N}\sum_\phi \frac{\sin(\phi N)\sin(\phi)
\tanh[\omega_\phi/T]}{\omega_\phi}
\end{eqnarray}

We show the correlation entropy $S(i:j)$ in the ground state as a function of
coupling $\lambda$ and distance between two site $r=i-j$ in Fig.
\ref{figure_centropy_00}. The result is also impressive. As is well known
\cite{Sachdev}, the ground state of the transverse field Ising model consists of
two different phases, whose corresponding physical picture can be understood
from both weak and strong coupling limit. If $\lambda\rightarrow 0$, all spins
are polarized along $z$ direction, the ground state then is a paramagnet and in
the absence of long-range correlation, while in the limit $\lambda \gg 1$, the
strong Ising coupling introduce magnetic long-range correlation in the order
parameter $\sigma^x$ to the ground state. The competition between these two
different order leads to a quantum phase transition at the critical point
$\lambda_c=1$. From Fig. \ref{figure_centropy_00}, we can see that the
correlation entropy tends to zero quickly as the distance between two sites
increases in the paramagnetic phase. This phenomena can be well understood from
the fact that the ground state in this phase is non-degenerate and almost fully
polarized, therefore the knowledge of the state at one site $i$ does not effect
the state of another site $j$ far away, which leads to zero information in
common between two sites. However, this scene is not true in another phase. When
$\lambda > 1$, the ground state is twofold degenerate and possess long-range
correlation. Before the measurement, the uncertainty of the state at an
arbitrary site is very large. However, if we learn it from one site, the state
at another site, even far away, is almost determined. Which leads to a finite
correlation entropy between two sites even if they are separated far away from
each other.

Obviously, the behavior of correlation entropy in the transverse field Ising
model is quite different from the quantum entanglement. In the previous works
\cite{AOsterloh2002,TJOsbornee} on the pairwise entanglement in the ground
state of this model, it has been shown that the concurrence vanishes unless the
two sites are at most next-nearest neighbors. In the paramagnetic phase, the
correlation entropy share similar properties in common with the concurrence. In
the ordered phase, however, the correlation entropy does not vanish even the
distance between two sites becomes very large, such as 50 lattice constant.
Moreover, the correlation entropy also shows interesting scaling behavior, just
as that of the concurrence, around the critical point, as is shown in Fig.
\ref{figure_centropy_scaling}. Moreover, we find that at the critical point the
first derivative of the correlation entropy between two neighboring sites
scales like
\begin{eqnarray}
\left.\frac{dS(0,1)}{d\lambda}\right|_{\lambda=\lambda_c}\simeq {\rm
const}\times \ln N.
\end{eqnarray}
or
\begin{eqnarray}
S(0,1)\simeq S(0, 1)|_{\lambda=\lambda_c} + {\rm const}\times
(\lambda-\lambda_c)\ln N.
\end{eqnarray}

However, for those sites are separated far away, the correlation entropy shows
quite different scaling behavior. For examples, in Fig.
\ref{figure_centropy_scaling_long}, we show the scaling behavior the
correlation entropy between two sites at the longest distance in a ring. This
first observation is that when $N\rightarrow\infty$, ${dS(0,N/2)}/{d\lambda}$
becomes divergent. Moreover, detailed analysis reveal that the maximum value of
the first derivative of the correlation entropy between two farthest sites in a
ring scales like
\begin{eqnarray}
\frac{dS(0,N/2)}{d\lambda}\simeq  {\rm const}\times \ln^3 N.
\end{eqnarray}
which differs from $\ln N$ for ${dS(0,1)}/{d\lambda}$. Obviously, these
interesting scaling behavior enable us to learn the physics of real infinite
system from the scaling analysis.

\begin{figure}
\includegraphics[width=8cm]{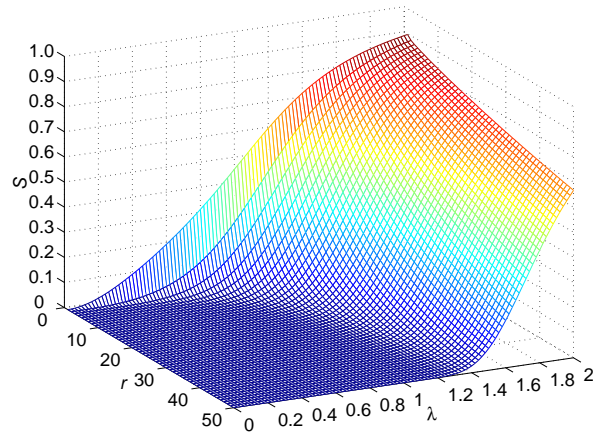}
\caption{\label{figure_centropy_02}(color online) The correlation entropy as a
function of $\lambda$ and the distance $r$ (in unit of lattice constant) at
$T=0.2$. Obviously, at low temperature, the correlation entropy at long distance
is destroyed in the quantum critical region around $\lambda=1$.}
\end{figure}

\begin{figure}
\includegraphics[width=8cm]{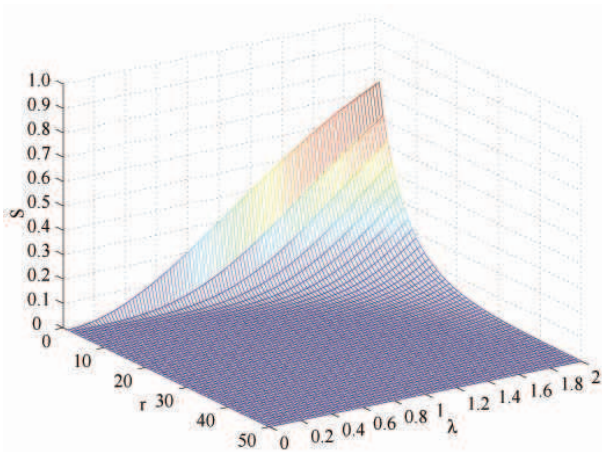}
\caption{\label{figure_centropy_05}(color online) The correlation entropy as a
function of $\lambda$ and the distance $r$ (in unit of lattice constant) at
$T=0.5T$. At higher temperature, the correlation entropy at long distance is
completely destroyed.}
\end{figure}

\begin{figure}
\includegraphics[width=8cm]{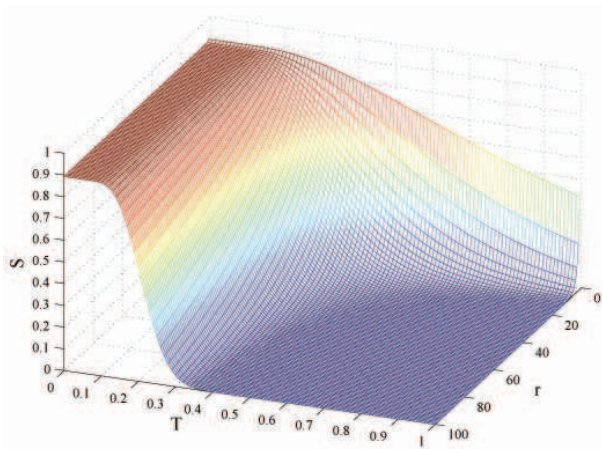}
\caption{\label{figure_centropy_l20}(color online) The correlation entropy as a
function of temperature $T$ and the distance $r$ (in unit of lattice constant)
at $\lambda=2.0$. We can see from the figure that the correlation entropy is
protected by an energy gap at low temperature. }
\end{figure}

On the other hand, there is no thermal phase transition in one-dimensional
quantum spin system according to the Mermin-Wagner theorem \cite{NDMermin66}.
Thus let us study the properties of the correlation entropy away from zero
temperature. The results are shown in Fig. \ref{figure_centropy_02} and
\ref{figure_centropy_05} for $T=0.2$ and $0.5$ respectively. We can see that at
lower temperature $T=0.2$, the correlation entropy is broken only around the
critical region $\lambda\sim 1$. As the temperature increases, the correlation
entropy in larger $\lambda$ region is also destroyed. These observations tell
us that the correlation entropy is a decreasing function of the temperature, as
we can see from Fig. \ref{figure_centropy_l20}. Since the broken symmetry only
exists in the ground state of an infinite system and the thermal fluctuation
tends to destroy the correlation entropy, it is not possible to reestablish
such a long-range correlation at finite temperatures. Therefore, no thermal
phase transition happens in the one-dimensional transverse field Ising model.
This observation, based on the numerical calculation, is consistent with the
Mermin-Wagner theorem.

\section{discussions and prospects}
\label{sec:dis}

With analytical studies and numerical calculations, we have discovered the
rigorous relation between the correlation entropy and the long-range
correlation. It strongly indicates us the non-trivial role of the correlation
entropy in the critical phenomena. This discovery motivates us to study both
the thermal and quantum phase transitions from the point view of information
theory, i.e. mutual information, whose non-vanishing behavior at long distance
really witnesses the violation of the extensive properties of the entropy in
the statistical physics, i.e. $S($AB$)=S($A$)+S($B$)$. In the two models we
studied in this model, since the correlation length at the critical point
diverges, the physics of the system has a strong dependence on the system size,
i.e. the scaling behavior. On the other hand, it also implies that the entropy
at the critical point is no longer a linear function of the volume of the
system, i.e. $S\neq sV$. In one dimensional system, it has already been noted
that entropy of the subsystem satisfies $S(x)\propto\ln(x)$ \cite{korepin2004}
where $x$ is the length of subsystem in the critical region of some spin
systems. Then, from this point of view, the spatial degree of freedom of the
system is suppressed around the critical point. This observation reminds us a
well-known holographic principle on the entropy of the black hole which says
that the entropy of the black hole is proportional to the surface. Though a
rigorous prove is still not available, we are sure there there are something in
common between the entropy in the critical phenomena and that in the black
hole, and this relation deserve for further investigation.

On the other hand, though we restrict ourselves to the two-point correlation
entropy in the above studies, if the system processes block-block order, such
as dimer order \cite{CKMajumdar69}, it may be useful to investigate the
properties of block-block correlation entropy. A simple example is
Majumdar-Ghosh model with the Hamiltonian ${H} = \sum_i \left( J_1 {\bf
\sigma}_i\cdot {\bf \sigma}_{i+1} + J_2 {\bf \sigma}_i\cdot {\bf \sigma}_{i+2}
\right)$, where $J_2$ is the coupling between two next-nearest neighbor sites.
In this model, if $J_2=1/2$, the ground state is a uniformly weighted
superposition of the two nearest-neighbor valence bond state
\cite{CKMajumdar69}:
\begin{eqnarray}
&&|\psi_1\rangle=[1,2][3,4]\cdots[L-1,L]\nonumber \\
&&|\psi_2\rangle=[L,1][2,3]\cdots[L-2,L-1]
\end{eqnarray}
where
\begin{eqnarray}
[i,j]=\frac{1}{\sqrt{2}}(|\uparrow\rangle_i|\downarrow\rangle_j-|\downarrow\rangle_i|\uparrow\rangle_j).
\end{eqnarray}
Then the block-block correlation entropy can help us to understand such a dimer
order.

Moreover, from the definition of our correlation entropy, it is also useful to
introduce a characteristic length for the statistical system, below which the
extensive properties of the entropy is violated. Such a characteristic length
has a non-trivial meaning, since the extensive property of the entropy in the
statistical physics is only valid above this scale. A simple example is the
molecule which is composed of some atoms. When we study the physics of molecule
gas, we have to regard a molecule as a whole because of its internal order.
Only when the temperature is high enough to break its order, the atom then
plays the important role to the statistical properties of the system.

Though we verify the non-trivial behavior of the correlation entropy in terms
of spin systems, the rigorous relation between the non-vanishing correlation
entropy and long-rang correlation is valid for all many-body systems.
Therefore, it is expected to provide more physical intuition into the critical
phenomena, such as Bose-Einstein condensation and superconductivity from the
point view of the correlation entropy. Take the former as a simple example, the
reduced density matrix between two sites in the space-time can also be
expressed in terms of bosons operators $a_{\bf x,
p}^\dagger|0\rangle=\exp(-i{\bf px})$ in space representation. At high
temperatures, $\langle{\bf p}|\rho_{\bf xx'}|{\bf p'}\rangle$ vanishes as ${\bf
x-x'}$ increases. Only when the condensation happens, $\langle{\bf 0}|\rho_{\bf
xx'}|{\bf 0}\rangle$ leads to a non-vanishing correlation entropy, just as what
we have shown for the correlation entropy in quantum spin system.

\section{Summary and acknowledgement}
\label{sec:sum}

In summary, the correlation entropy plays a universal role in understanding
critical phenomena. Its non-vanishing behavior not only help us have deep
understanding to the entropy in the statistical physics, but also shed light on
the long-range correlation in the critical behavior.

This work is supported by the Earmarked Grant for Research from the Research
Grants Council of HKSAR, China (Project CUHK N\_CUHK204/05 and HKU\_3/05C) and
UGC of CUHK; and NSFC with Grants No. 90203018, No. 10474104 and No. 60433050.
We thank Kerson Huang for the helpful discussion.

\end{document}